# Evolution of the energy consumed by street lighting in Spain estimated with DMSP-OLS data

Alejandro Sánchez de Miguel,[1] Jaime Zamorano,[1] José Gómez Castaño[1] and Sergio Pascual[1]

[1] *Dept. Astrofísica y CC. de la Atmósfera. Universidad Complutense de Madrid, Avda. Complutense s/n, Madrid, 28040 Spain*

*tel: +34 91 394 52 38, fax: +34 91 394 4635, e-mail: alejasan@ucm.es*

**Abstract**

We present the results of the analysis of satellite imagery to study light pollution in Spain. Both calibrated and non-calibrated DMSP-OLS images were used. We describe the method to scale the non-calibrated DMSP-OLS images which allows us to use differential photometry techniques in order to study the evolution of the light pollution. Population data and DMSP-OLS satellite calibrated images for the year 2006 were compared to test the reliability of official statistics in public lighting consumption. We found a relationship between the population and the energy consumption which is valid for several regions. Finally the true evolution of the electricity consumption for street lighting in Spain from 1992 to 2010 was derived, it have been doubled in the last 18 years in most of the provinces.

## 1 Introduction

Satellite images taken during the night have changed our image of the Earth and they are now one of the most efficient tools in the fight against light pollution. Many studies on the relationship between the amount of light registered by satellites, as DMSP with OLS, with parameters as population density and distribution, power consumption, and economic activities, to name but a few, have been published ([10][11][12][13][14][15][37][23][46][47][48]).

The relationship between the light pollution and the emission registered from satellites has been treated by Cinzano et al. in several papers ([3][4][5][6][7][8]). The main contributor to light pollution is public street lighting ([14] [16] [22]). It is for this reason that our group has worked on this area since 2007 with the aim of finding a procedure to estimate energy power consumption in public electric lighting using both statistics and satellite data ([3][25][26][27][28][29]). This paper improves on the previous findings and analyses by explaining the specific method to intercalibrate images and to compare these data with the statistics.

We outline our method and results for Spain. Our goal is firstly to determine and to compare energy efficiency for states of the European Union (EU) and then for the rest of the world in a second phase. When colour data of the light is available (with spectral data or at least with several photometric bands), information of the emitting sources can be obtained and therefore it is necessary to have more accurate information about the real energy consumption and efficiency ([14], [18]).

Our method is based on empirical calibration of satellite data using statistics on the power consumption. The total power consumption is readily available but the budget on the electric power for public lighting is hard to find. We have only had access to the VITO report ([32]), to some sparse data for Italy ([24]), and to the official data from Spanish Ministry of Industry. Using statistics from the year 2007 from the

Ministry of Industry, we established a relationship between the flux observed in the calibrated images of the DMSP-OLS satellite for the year 2006. Following this, a calibration method was used to establish a relationship between DMSP-OLS images and the power consumption to verify its evolution. Once the method was validated using the statistics, we compared the results with the data of the Ministry between the years of 1975 and 2007.

This procedure allowed us to estimate the power consumption for street lighting from 1992 to 2012 based on satellite imagery. Our analysis is purely statistical, despite the obvious relationship of physics between the electricity consumption with the emission produced by the light sources detected in the satellite images. However, a complete physical model should take into account factors such as the type of emission source, which can increase the estimation by 100%, in the consumption of specific regions. Other atmospheric effects, such as scattering, have not been taken into consideration because it is a second order effect which does not dominate on a large scale. As it is evident from the development of this work, the radiance seems to be directly correlated with the consumption and therefore higher order corrections are not needed.

## 2 Data

### 2.1 DMPS-OLS images:

Our main source of satellite data is the radiance calibrated DMSP-OLS images corresponding to 1996-1997 ([13]) and 2006 ([36]). There are other images calibrated but not accessible to the public. There were non-calibrated DMSP-OLS images of 1992-2010 ([1], [10], [13] & [15]). Our future work will include data from Suomi NPP/VIIRS [41]. This data is provided by the Earth Observation Group (NOAA National Geophysical Data Center). An additional source of information is the nocturnal pictures taken by astronauts aboard the International Space Station (ISS) [42]. These images taken with digital cameras that have three color channels and can be calibrated ([30], [34]). The data extraction was performed using GIS techniques, unlike in [29], where it was performed a manual extraction.

### 2.2 Available statistics data:

The Spanish Ministry of Industry, Energy and Tourism webpage publishes annual reports ("Estadísticas Eléctricas Anuales") [40][1] which provide information for each activity sector. National data from 1958 to 2007 are available, whereas they provide information for each Spanish province from 1975.

As we have already shown in previous work, there are some discrepancies between the official data and the true power consumption (see Figure 6 of [26]). Some provinces show abnormally decreasing values between 1985 and 1992 and the national power consumption seems to be stabilized before an increase from 1992 onwards. Using official data from the book "La Energía en España 2004" [43] and from information from the years 1936 to 1957, collected at the Instituto Nacional de Estadística (INE), we have rebuilt the real evolution of the electric power consumption in public lighting for Spain [27]. An exhaustive analysis of the statistics was carried out to find the source of these errors. The official data, provided by the Spanish Ministry, have already corrected this error, however only for the year 2007 as well as some less populated provinces for all the years as it is easy to see in Figure 10.

A test was conducted on the official correction of the data. We performed a comparison between energy consumption and the cumulative radiance for 2007 following the prescriptions of [13]. Figure one shows the line of best fit for the data. The fit is very good, based on the Pearson correlation coefficient of $R^2=0.93$. All the analyses have been done with Python scripts.

---

[1] http://www.minetur.gob.es/energia/balances/Publicaciones/ElectricasAnuales/Paginas/ElectricasAnuales.aspx

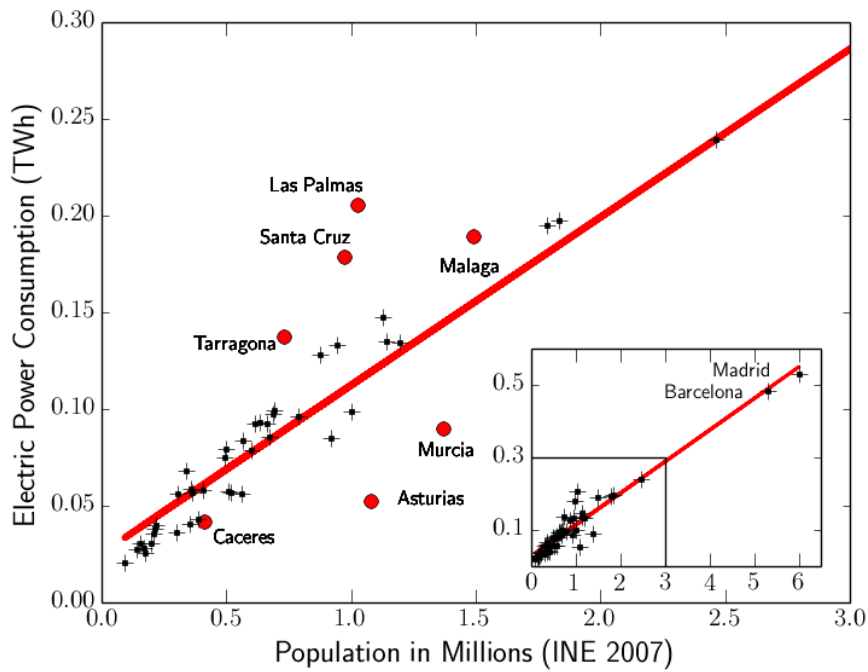

Figure 1. Power consumption in street lighting versus population.

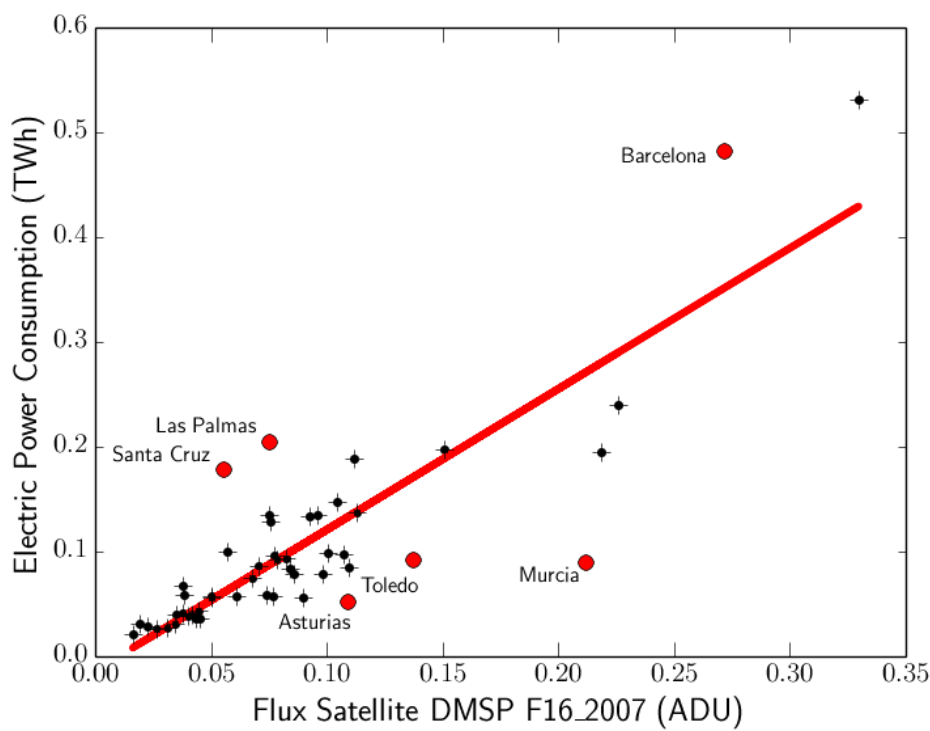

Figure 2. Electric Power Consumption in street lighting versus Flux F16_2007.
Madrid and Barcelona, top right, are clearly saturated.

Some provinces that do not follow the relationship of Figure 1 are also anomalies in Figure 2. It seems that they have an excess of the measured flux from satellite or from power consumption. It is probable that the statistics for these provinces are not accurate, as can be seen in the evolution of some provinces (for example see blue points in Figure 3) or check [27].

The most populated provinces (Madrid, Barcelona or Valencia) have additional data available. For the region of Madrid we have analyzed the statistics for Comunidad de Madrid[2] (the whole region) and (Madrid City).[3] We have been able to rebuild the true evolution of power consumption (see Figure 3) using all the available data on 'official lighting', 'public lighting' and 'other types of lighting' by taking into account the different administration rules to archive power consumption data. The statistics for the Madrid region (data taken from the Ministry) only includes power consumption for Comunidad de Madrid between 1986 and 2005 and to Madrid City. Data for Madrid City is divided into neighborhoods, lamp types and the number of luminaries since the year 1975.

## 3 DMSP-OLS visible sources

Official statistics show that the electric power consumption increases in an upward trend. We expect light emission to vary in the same way. Most of the sources detected from space in the 0.3-0.9 microns range are public street lighting. Other sources of light are: (1) Ornamental or security lights (city centers, monuments, airports, open-air mineral extractions, commercial, sports, greenhouses) [22] and (2) Oil extractions, fires and boats. These transients are mostly removed by [1] in the DMSP-OLS Stable lights products.

---

2   http://www.madrid.org/iestadis/fijas/estructu/general/anuario/descarga/anu12-2-3.xls
3   http://www-2.munimadrid.es/CSE6/jsps/menuBancoDatos.jsp

Light sources for group (1) are metal-halide lamps in Spain and they can be easily found in ISS images mainly in city centers. They are usually excluded from statistics because they are not public street lighting, as they appear saturated and their contribution varies during the night (Sanchez de Miguel et al. in prep.) & [22]. The Flaring sources are that uncommon in Spain that they do not appear in the Global Gal Flaring Estimates of the Earth Observatory NOAA [44].

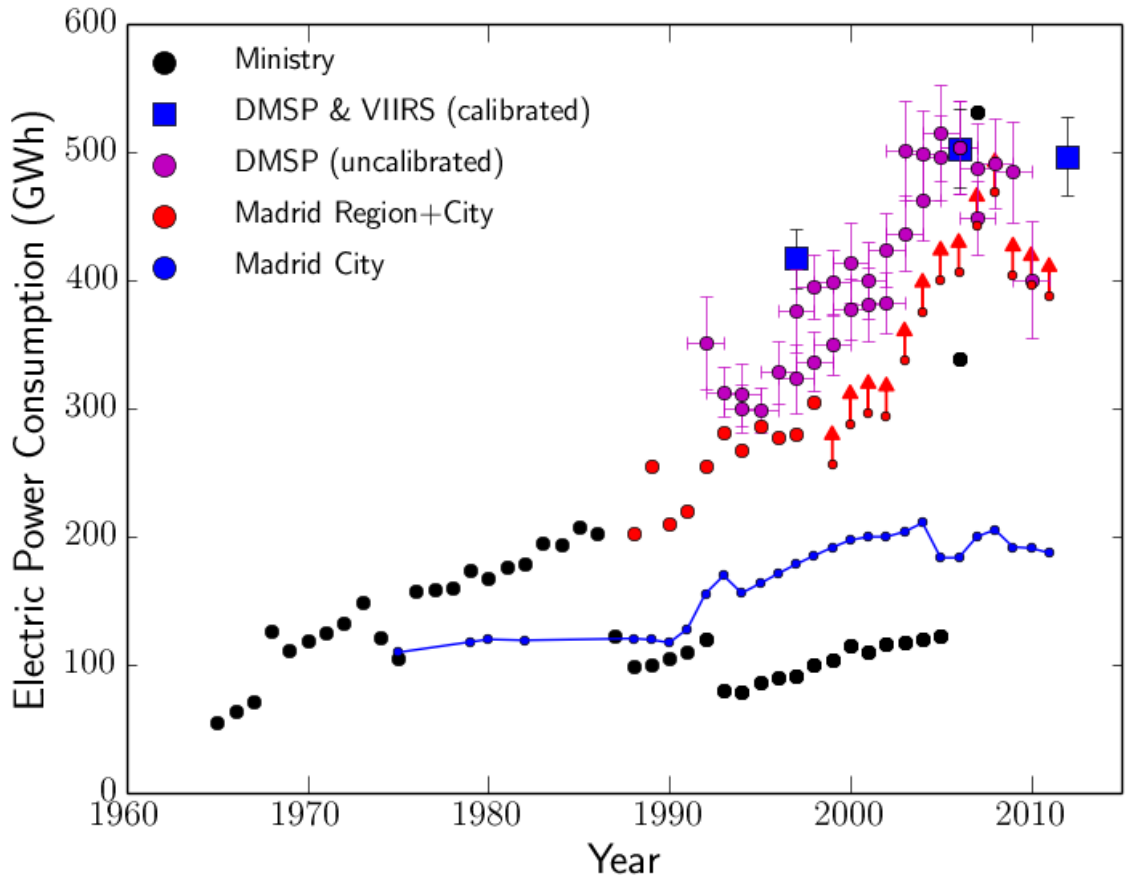

Figure 3: Madrid region electric power consumption for street lighting. The data from the Ministry is in black, the data from Comunidad de Madrid is in red, as the original data of Comunidad de Madrid had not Madrid city included between 1986 to 1999 they had been added, then it's shown the original data of the Comunidad de Madrid with year with arrows because they don't match with previous data. Madrid city data is in blue dots with blue line. Blue squares, estimated consumption using radiance calibrated data, DMSP 1997, DMSP 2006 and testing 2012 VIIRS. The magenta dots came from the non-calibrated images. The blue square's error bars have been calculated using the 90% confidence interval of the Power consumption vs. radiance fit. The magenta error bars have been calculated using the satellite dispersion correction and the 90% confidence interval to represent power consumption vs. radiance fit (see additional e-components for detail figure of each province).

Light sources of group (2) are also out of the official statistics but not being stable, they are not present in the final calibrated satellite DMSP/OLS images or they are only present in the 50% ([12]). They present spectra of black bodies of different temperatures.

## 4 Electric power consumption using satellite images

### 4.1 Empirical calibration using radiance calibrated satellite images

We have obtained the cumulative radiance for the Spanish provinces following the methodology described by [13]. In comparison to the power consumption, we do not use the total consumption, but the public street lighting, more related to the light emission detected by DMSP satellite. It should be kept in mind that the amount of light also depends on many other parameters such as the ground reflectivity, the lamp efficiency and the street design to name but a few. We have already shown in Figure 2 the relationship, even using the non-calibrated image.

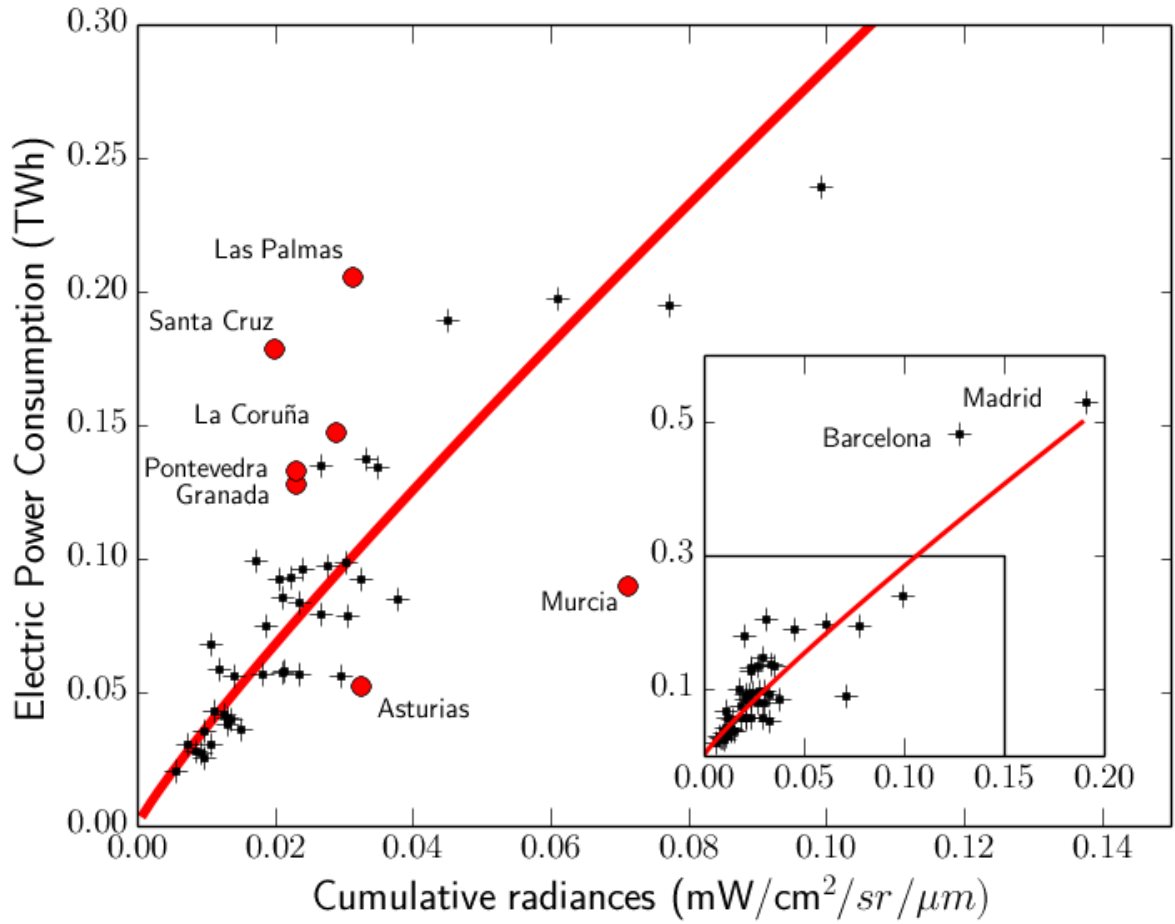

Figure 4: Fit to the power consumption in street lighting to cumulative radiances. Potential fit to the data is shown in red. The inset shows an extended plot to include Madrid and Barcelona.

The relationship between the power consumption in street lighting and the calibrated cumulative radiance is plotted in Figure 4. To reject outliers we use a robust linear fit as explained in [33]. Next fit is potential $\log 10(y) = m \log 10(x) + c$.

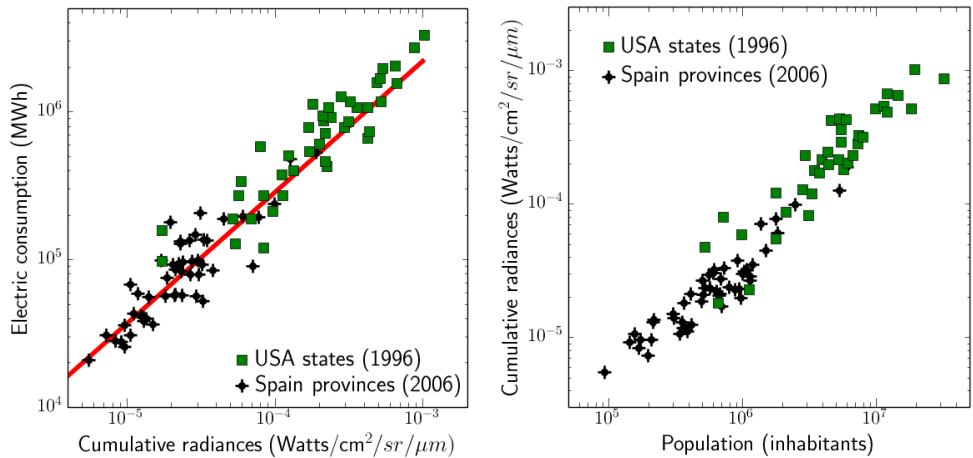

Figure 5:(Left) Comparison between the total power Consumption in Elvidge et al. 1999 dived by factor 80, and the street light power consumption in Spain. The same lines in Figure 4.

Figure 6: (Right) Population versus Cumulative radiance. USA by Elvidge et al. 1999, Spain this work.

To compare with previous pioneering works in this field we represent our data in Figure 5 and compare it with that of the states in the USA from [13]. The power consumption for the states of the USA has been divided by 80 to estimate the consumption in public lighting (around 1.77% [38], although the typical lighting efficiency in USA's lamps are 84 W/lm [38] against the 121 W/lm in Spain [32].) So, we have corrected this offset and the continuous line (fit Figure 4) is compatible with these data. Area correction (see [15], [19]) is not needed in this case as the differences for Iberian Peninsula are less than 10% and the latitude is similar to USA.

## 4.2    Cross calibration of DMSP-OLS satellites

Radiance calibrated data is the key ingredient to study the evolution of electric power consumption in public lighting from satellite data. Some efforts have been made in the past to cross-calibrate DMSP-OLS data since there is an evident relationship between calibrated and non-calibrated images as shown for instance in [13],[48] or in Falchi private communication 2009. Our strategy consists of using data for Spanish provinces where there are almost no light sources of group #2 but many of group #1, mainly found in city centers. It is interesting to note that the ornamental lights are variable (they are switched off during the night) and they appear only if there are present in more than 50% of the DMSP-OLS images [1]. Pictures taken by astronauts from ISS clearly show this effect when images at different hours during the night are compared.

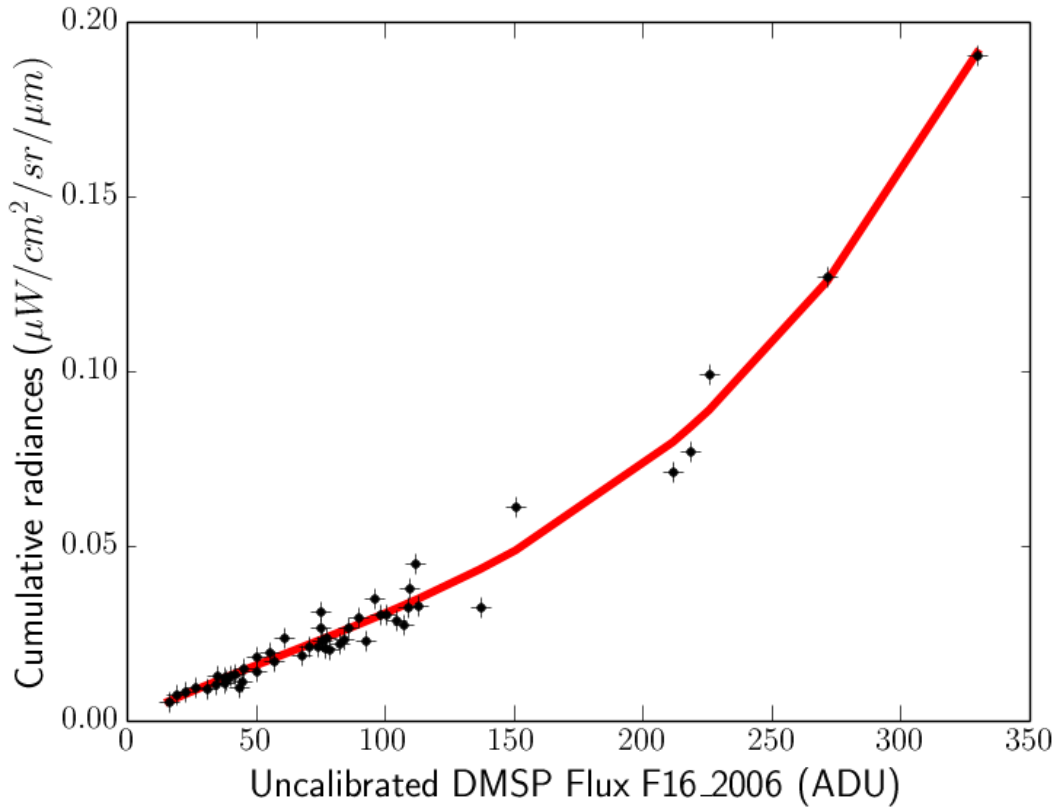

Figure 7: Relationship between the non-calibrated F16_2006 and calibrated images for Spanish provinces. Red line is a polynomial fit of order 4

A polynomial fit of 4th order is enough to fit the cumulative radiances for Spanish provinces using calibrated data [36] and the flux in ADU of DMSP-OLS F16_2006 (see Figure 7). Elvidge et. al. 1999 [13] explains how darker zones appear brighter due the diffuse lighting that originates in bright and saturated areas [45]. We are working on this problem using nocturnal ISS pictures taken in HDR mode with increasing exposure times. It seems that this effect is real and not an artifact or an instrumental problem ([30], [31], [34], [45], [47] & [49]).

There are systematic differences between DMSP satellites and over the years as many authors have demonstrated (see Figure 14 of Falchi 2005 at [7], [15], [19] & [23]). This is why it is not straightforward to apply this fit to all satellite data. To resolve this problem [15], it is necessary to analyze one region point by point. But there are differences between regions and we prefer to assume, as a first step, that there is a calibration for each region. If the results are consistent, the calibration could be used for any region in the world.

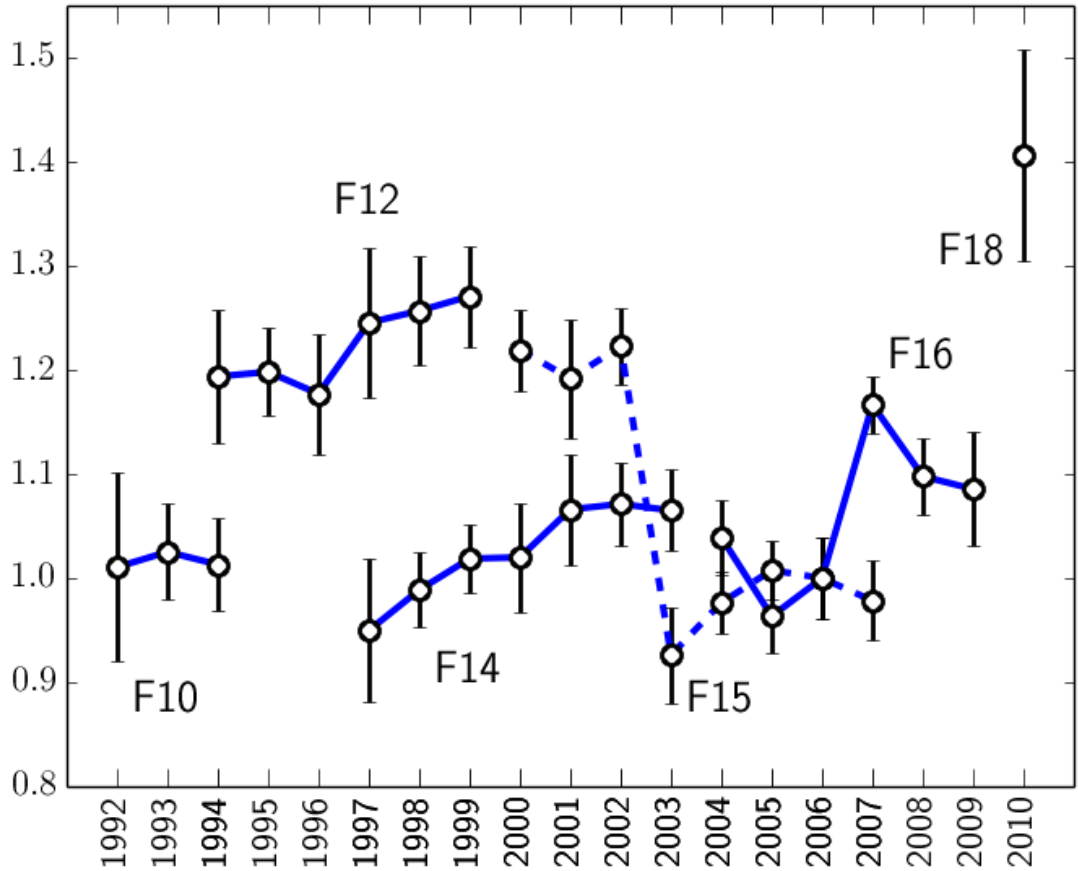

Figure 8: Offsets between images and satellites obtained following the procedure explained in figure 9. The F16_2006 has been set as reference. The error bar indicates the standard variation of the relative values of each region to the expected using the interpolation of DMSP calibrated images 1996-1997 and 2006.

Assuming that there is a smooth evolution in the light emission to space, based on the smooth evolution on the best data of evolution of energy consumption. We are able to use a linear fit for calibrated data for F12_1997 & F16_2006 and obtain a first estimate of the expected values for cumulative radiances for the years between 1992 and 2010 of each of region (up to 50 Spanish provinces). The residuals show the offset for each satellite/year with different dispersions that we consider a random error in flux. Thus we take F16_2006 as reference and use the mean values of these offsets to correct all the satellite data. The estimated cumulative radiance for each region is obtained using this empirical calibration on the non-calibrated data. This procedure reduces data dispersion whilst keeping the evolution of each region. All entire procedure is summarized graphically in Figure 9.

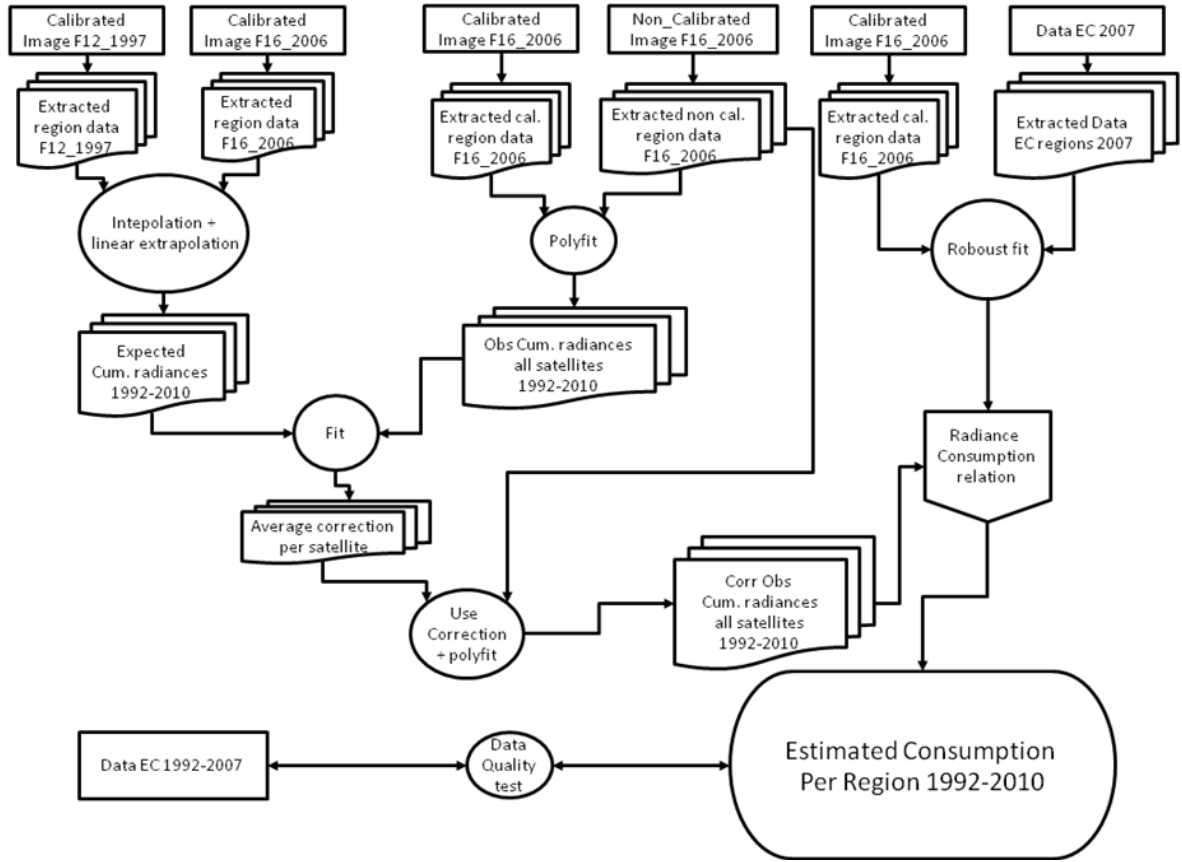

Figure 9: Calibration procedure. Right branch top: Relationship between the flux from the integrated regions in the calibrated image F16_2006 and Energy consumption of 2007 per region. Centre branch top, Relationship between integrated flux in calibrated image F16_2006 and non-calibrated F16_2006. Then apply this relationship to all non-calibrated data. Left branch top: Estimated grow between 1992-2010 using Calibrated data 1996-7 and 2006. Fit Centre Branch with left branch and correct the systematic offset per satellite. Correct original non-calibrated with the offset from the satellites. Use the relationship between non-calibrated and calibrated 2006. Then use right branch data (relationship between energy consumption and calibrated data) to obtain an estimation of the energy that had been consumed. Then Check with the stats of energy consumption on street lighting.

**4.3   Calibration of Electric power consumption in public lighting versus cumulative radiances**

The quality of our empirical calibration we tested by comparing the evolution of the cumulative radiances for the respective region studied with the statistics in electric power consumption in public lighting as explained above. The panels of Figure 10 show the plot for each one of the Spanish provinces.

The correlations are extremely good for the less populated provinces (i.e., Ávila, Cantabria, Burgos, Huesca, Soria …) whose official statistics seems to be accurate. The inconsistencies found for the most populated provinces, where power consumption has only been similar to that expected from satellite data in recent years, are due to the change in electric lighting statistics which was mentioned above for the case of Madrid.

For some provinces it seems the shape of evolution is correct, with an absolute difference. The differences could be due to poor empirical calibration, however other explanations are possible as they fail to include some cities in the total power consumption statistics or use luminaries with different efficiencies. We are working on a detailed study for the region of Madrid, where detailed official statistics data are available, using RGB color data from ISS nocturnal pictures, SUOMI-NPP/VIIRS and DMPS-OLS.

These good agreements between the DMSP-OLS radiances and public power consumption in street lighting do not necessary imply that all emissions came from this type of sources. In fact, a recent study found that only 44-55% of consumption maybe considered as public street lighting (streets, parks, motorways …) [22]. This issue will be studied for the case of Spain and will be discussed in future studies, but this statistical relationship implies that the same scale factor is applied to public lighting and private lighting.

# 5  Conclusions and results

We have shown whown power consumption data relates to calibrated and non-calibrated radiance images. There is still a considerable uncertainty for high radiance regions where there are many sources of error which cannot be considered as stable sources, in fact it is in these regions where light pollution may come from other sources not related to public lighting. In spite of this, and the fact that these regions are saturated by non-calibrated images, they can be measured thanks to their ability to produce more diffuse light on not lit regions (overglow [45]). This diffuse light comes from point spared function of the instrument and the atmospherically diffusion. The effect of projection has not been taken into account in order to make a correction on the area represented by each pixel in order to compare with the work of [13]. Having said this, an analysis for Europe is underway considering this effect without moving the information as in [2] but using reprojection [15]. The most populated provinces appear to have begun to stabilize the growth of expenditure on public lighting (Figure 10). However this does not occur in the less populated provinces which continue to grow at a similar rate in spite of the economic crisis. The general trend of the country had been grow nearly constantly by 18 years as we can see in Figure 11. Only in the last year maybe the country had stop growing in power consumption.

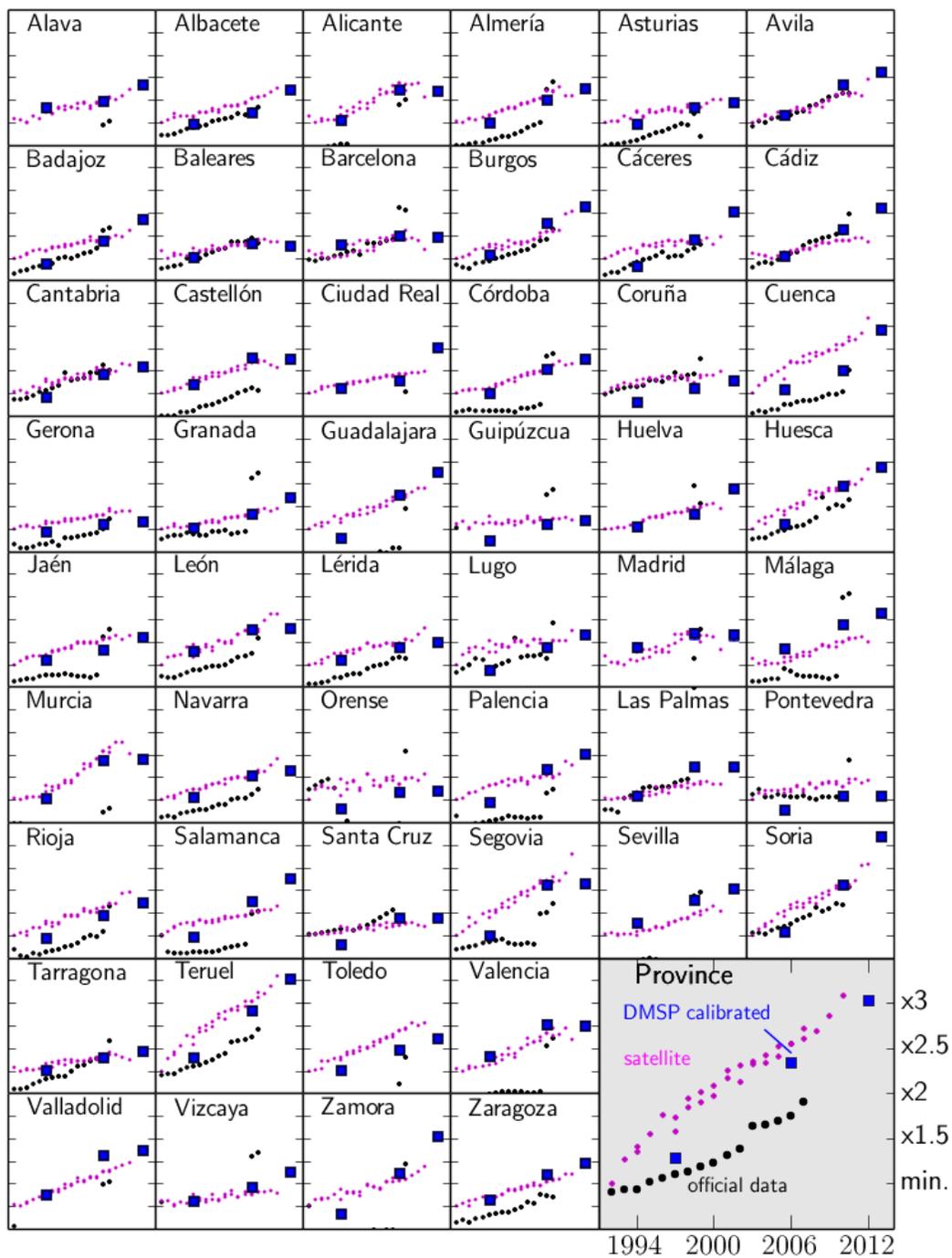

Figure 10: Evolution of power consumption in street lighting. The official data is in black dots. The non-calibrated data in magenta and the calibrated data in the form of blue squares. 1996-1997 DMSP-OLS, 2006 DMSP-OLS and testing 2012 SUOMI-NPP/VIIRS. Scale are different for each plot. Limits in power consumption have been set as plotted in the lower right panel to show variation in each province.

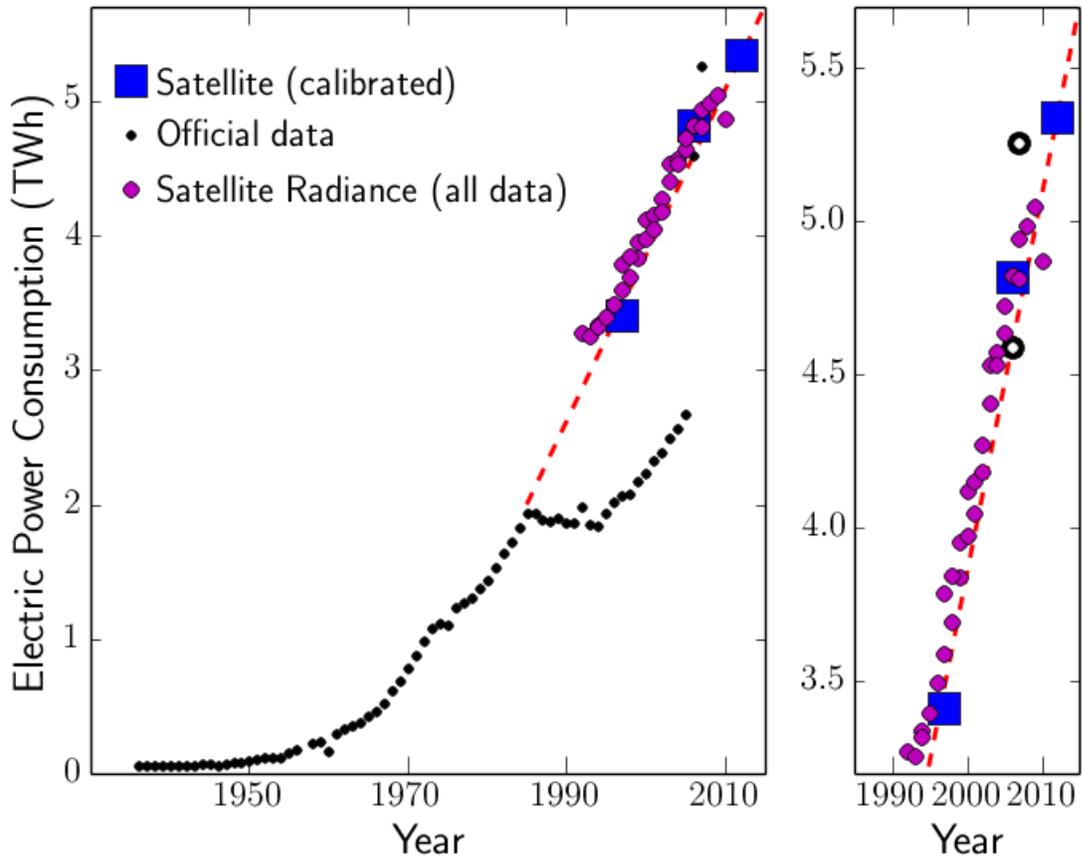

Figure 11: Evolution of power consumption in street lighting in Spain. Official data, black dots. Magenta dots, non-calibrated data DMSP. Blue squares, 1996-1997 DMSP, 2006 DMSP, testing VIIRS. The detailed view of the last years (inset plot) shows the correction of the official data after our works [26], [27].

**Acknowledgments**

This work has been partially funded by the Spanish MICINN (AYA2009-10368, AYA2012-30717), by the Spanish programme of International Campus of Excellence Moncloa (CEI) and by the Madrid Regional Government through the AstroMadrid Project (CAM S2009/ESP-1496). Thanks goes also to Francisco Ocaña and Jessica Starkey for the critical review of this text.

*Evolution of street lighting in Spain, Sánchez de Miguel*

*Accepted in JQSRT 26/11/2013 – Light pollution special Issue*                                                                 *17*